\documentstyle[epsfig,epsf,rotate,a4,12pt]{article}
\epsfverbosetrue

\def\slop{\rm{\Phi _{SL}}}
\def\flop{\rm{\Phi _{FL}}}
\def\pstot{\rm{\Phi _{Tot}}}

\begin{document}

\hfill   {\bf AS-TEXONO/05-06}\\

\begin{flushleft}
{\bf \Large
\underline{WCU/UI'05 Conference, Beijing, August 2005}}\\
{\bf Session : Cavitation Field and Sonoluminescence (3)}\\
{\bf Paper No. : ThBam2-06}\\
{\bf Date :  \today} \\
\end{flushleft}

\begin{center}
\Large
\bf  
{
Observation of Fluorescence Emissions from\\
Single-Bubble Sonoluminescence\\
in Water doped with Quinine
}
\end{center}

\begin{center}
\large
J.Q.~Lu~$^{\alpha , \beta}$,
H.T.~Wong~$^{\alpha ,}$\footnote{Corresponding Author $-$ 
Email: htwong@phys.sinica.edu.tw;
Tel.: 886-2-2789-6789;
FAX: 886-2-2788-9828.}
F.K.~Lin~$^{\alpha}$,
Y.H.~Liu~$^{\alpha , \gamma}$\\
\end{center}

\normalsize

\begin{flushleft}
$^{\alpha}$ Institute of Physics, Academia Sinica, Taipei 11529, Taiwan\\
$^{\beta}$ Department of Physics, Nanjing University, Nanjing 210093, China\\
$^{\gamma}$  Department of Nuclear Physics,
China Institute of Atomic Energy, Beijing 102413, China\\
\end{flushleft}
\begin{center}
\end{center}


\newpage

\normalsize

\begin{center}
\large
{\bf Abstract} \\
\end{center}
\normalsize
Sonoluminescence is a phenomenon involving
the transduction of sound into light.
The detailed mechanism as well as the energy-focusing
potentials are not yet fully explored and understood.
So far only optical photons are observed,
while emissions in the ultra-violet range are
only inferred.
By doping the fluorescent dye quinine into water
with dilute sulphuric acid,
the high energy photons can be converted into
the optical photons with slower decay constants.
These sonoluminescence and 
fluorescent emissions were observed in coincidence,
and the emitted signals of the two modes can be 
differentiated by their respective timing profiles.
Plans for using this technique as a diagnostic
tool to quantitatively study ultra-violet and other
high energy emissions in
sonoluminescence are discussed.

\begin{flushleft}
\end{flushleft}

\newpage

\section{Introduction}

Sonoluminescence (SL) are flashes of light emissions
due to collapsing ultrasound-driven bubbles.
Stable Single-Bubble Sonoluminescence (SBSL)\cite{slreview,rmp02}
was discovered in 1989 and have been under active studies.
The detailed physics responsible for SL emissions as well
the potentials and limitations
for achieving very high temperatures
are not fully explored and understood.

The detectable emissions from SBSL are
so far restricted to
the optical range (200~nm to 700~nm), the
only window in the electromagnetic spectrum where
water becomes transparent.
In addition, the typical sensitive range
for photo-multipliers (PMTs) commonly used
in SL detection is $>$300~nm and peaks at
$\sim$400~nm.
Spectral analysis\cite{spectral}
suggested a temperature of 25000~K at
the light-emitting surface if
the spectrum would follow that for black-body 
radiations. However, since only the tail of
the black-body spectrum is measure-able, 
there are substantial
uncertainties in the temperature derivation.
The same data in Ref.~\cite{spectral} was interpreted
by other author to indicate a temperature of
40000~K\cite{rmp02}.
In this case, the dominant energy fraction
of the SBSL emissions should be in the ultra-violet
(UV) range ($<$300~nm), such that most energy
from SBSL would be absorbed in the water and 
remains undetected. 
Still, the black-body model does not give the complete
description since it fails to
explain the lack of dependence of
the pulse width to wavelength\cite{rmp02,gompf97}.
In addition,
there are recent controversial 
claims of tritium and neutron emissions
from SL on deuterated-acetone\cite{ornl}
which, if confirmed,
would indicate that the nuclear reaction temperatures
of the order of 10$^7$~K can be achieved.
In addition, existence of plasma 
at the inner core of the collapsing bubbles  
in SBSL with sulphuric acid
was recently demonstrated\cite{plasma},
from which 
the derived temperature was 
8000$-$15000~K.
Another recent work\cite{slmhz} 
reported on SBSL emissions
at the 1~MHz range, where the 
observed optical spectrum  
implies internal bremsstrahlung  
from a transparent plasma core at
$\sim 10^6$~K.

It is therefore
important to devise methods to experimentally
probe the UV and other high energy 
emissions and to quantitatively
measure the photon energy emitted.
Measurements of the ratio
between UV to optical photon energy can 
provide a probe to the temperature of the
light emitting region.

This article describes the first step
in our research
efforts to address this question.
The key ingredient is to dope the
water with a fluorescent materials
such that the energy of the UV emissions
is converted (or ``wavelength-shifted'') 
to fluorescent light (FL)
in the optical range
which can then traverse the water to
be detected by PMTs.
Observations of such FL emissions in SBSL
were implicit in a previous study\cite{surfactant}
where the emphasis were on the effect
of active solutes to bubble dynamics and,
in particular, the emission spectra.
It was shown, averaging over a larger number
of events, that the spectra emitted by SBSL
in water doped with pyranine
is shifted to peak at green (520~nm).
The focus on this work is to study the
timing profiles of SBSL emissions 
in water doped with a fluorescent dye,
and to de-convolute the corresponding
SL and FL components.

\section{Experimental Set-Up}

\begin{figure}
\centerline{
\epsfig{figure=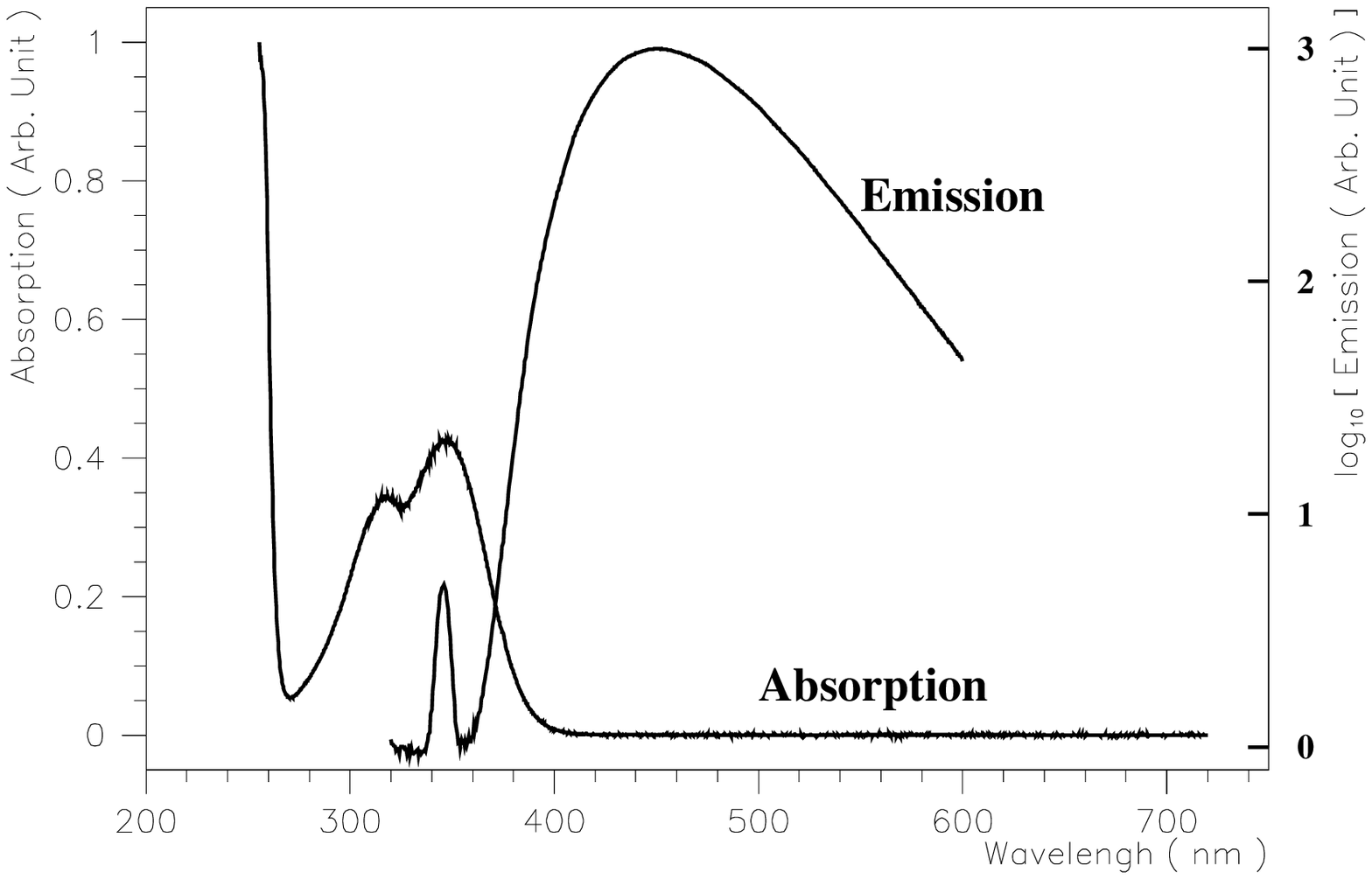,width=12cm}
}
\caption{
The emission and absorption spectra of
quinine-doped water with
dilute sulphuric acid, 
adapted from Ref.~\cite{quininespectra}.
The emission is due to an excitation
wavelength of 310~nm.
}
\label{spectra}
\end{figure}

Most commonly used fluorescent dye with large
light yields are not readily soluble in water or
inorganic solvent.
They are typically dissolved in  organic solvent
like xylene or pseudo-cumene to function
as liquid scintillators for the detection of
high energy radiations.
The fluorescent dye quinine 
(chemical formula: $\rm{ C_{20} H_{24} N_{2} O_{2} }$)
was selected for our investigations since it
readily dissolves in water mixed with
dilute acid.
The absorption and emission spectra of 
quinine\cite{quininespectra} are
depicted in Figure~\ref{spectra}. There is
little attenuation ($< 0.5 \%$) to the emission spectra
through 5~cm of water\cite{waterabsorption}.
In our measurements, 1~g of quinine was dissolved
in water mixed with 0.5 M(mole per liter) of 
dilute (that is, about 0.9\%) sulphuric acid. 
It has been checked that this small admixture
of sulphuric acid to water does not change
observably the light yields and pulse shapes
in SBSL, relative to those from pure water. 
Accordingly, we use ``water'' to 
denote 0.5~M dilute 
sulphuric acid for simplicity reasons
in this article.

\begin{figure}
\centerline{
\epsfig{figure=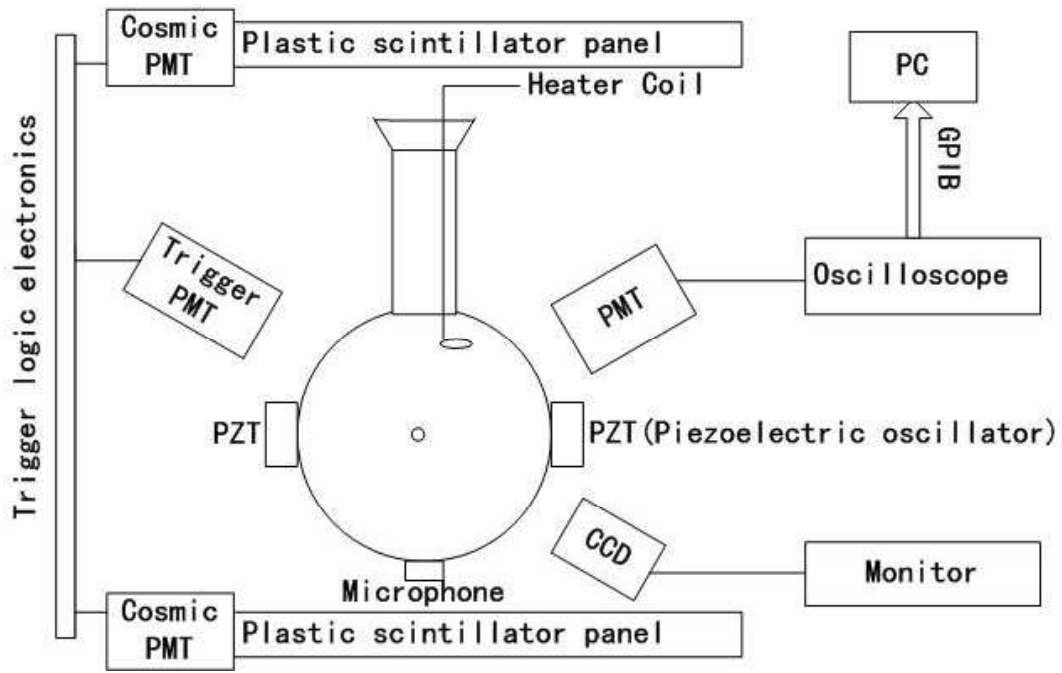,width=14cm}
}
\caption{
Schematic diagram of the experimental
apparatus, including the cosmic-ray
telescope.
}
\label{setup}
\end{figure}

The schematic diagram of the experimental
set-up is
displayed in Figure~\ref{setup}.
The standard SL configurations, gas manifold and
water processing procedures\cite{slreview}
were adopted.
The resonator was a spherical flask
6.4~cm in diameter,
driven by two piezoelectric (PZT)
oscillators at the resonance frequency of
about 25.5~kHz through a function generator
and power amplifier.
A third PZT attached to the flask
functioned as microphone monitoring
the shock waves generated during
bubble collapse.

Pure water was first degassed in a beaker inside
a vacuum chamber,
and a controlled gas composition was
introduced while the beaker was stirred
so that the water was in equilibrium
with the gas content, where
10\% air was used in this measurement. 
Sulphuric acid was added to the prepared water  
and quinine was subsequently dissolved into it.
The liquid was placed in the resonator flask
in a light-tight
refrigerator box with temperature
control between 0$^o$C and 20$^o$C.
The seed of the bubble was generated by a
NiCr heating wire.

The light emission
were detected by two PMTs
located directly outside the flask.
One PMT was used to provide the trigger,
while the signals from the other were sampled.
The rise time of the PMT is about 2~ns, while
the bialkaline photo-cathode
has a maximal
quantum efficiency at blue (420~nm).
Images from a CCD camera provided independent visual checks.
Output from the function generator, amplifier,
microphone and the PMT were all monitored
with oscilloscope to ensure stability and
that the proper phase relations were maintained.
Besides these standard set-up,
a cosmic-ray telescope (CRT) consisting
of two 7X6X3~cm$^3$ plastic scintillator
panels read out by PMTs 
were installed on top and bottom
of the resonator flask. Events triggered by
the CRT were used to study the characteristics
of FL emissions. 

The PMT pulse signal
were recorded via
a digital oscilloscope
at 2.5~GHz sampling rate. 
An event-by-event data acquisition
was performed, at a rate of about 5~Hz.
The data were transferred
from the oscilloscope via a GPIB-USB interface to
a PC using the commercial LabView
data acquisition software package.
The total light yield is proportional to
the integrated PMT signals.

\section{Observations of Fluorescence Emissions}
\label{sect::pulse}

Light emissions from SBSL 
in water and 
in quinine-doped water were measured.
The PMT signal profiles of individual events
were measured. The pulse shapes reported
in this Section are from the averaging over
1000 events.

\begin{figure}
\centerline{
\epsfig{figure=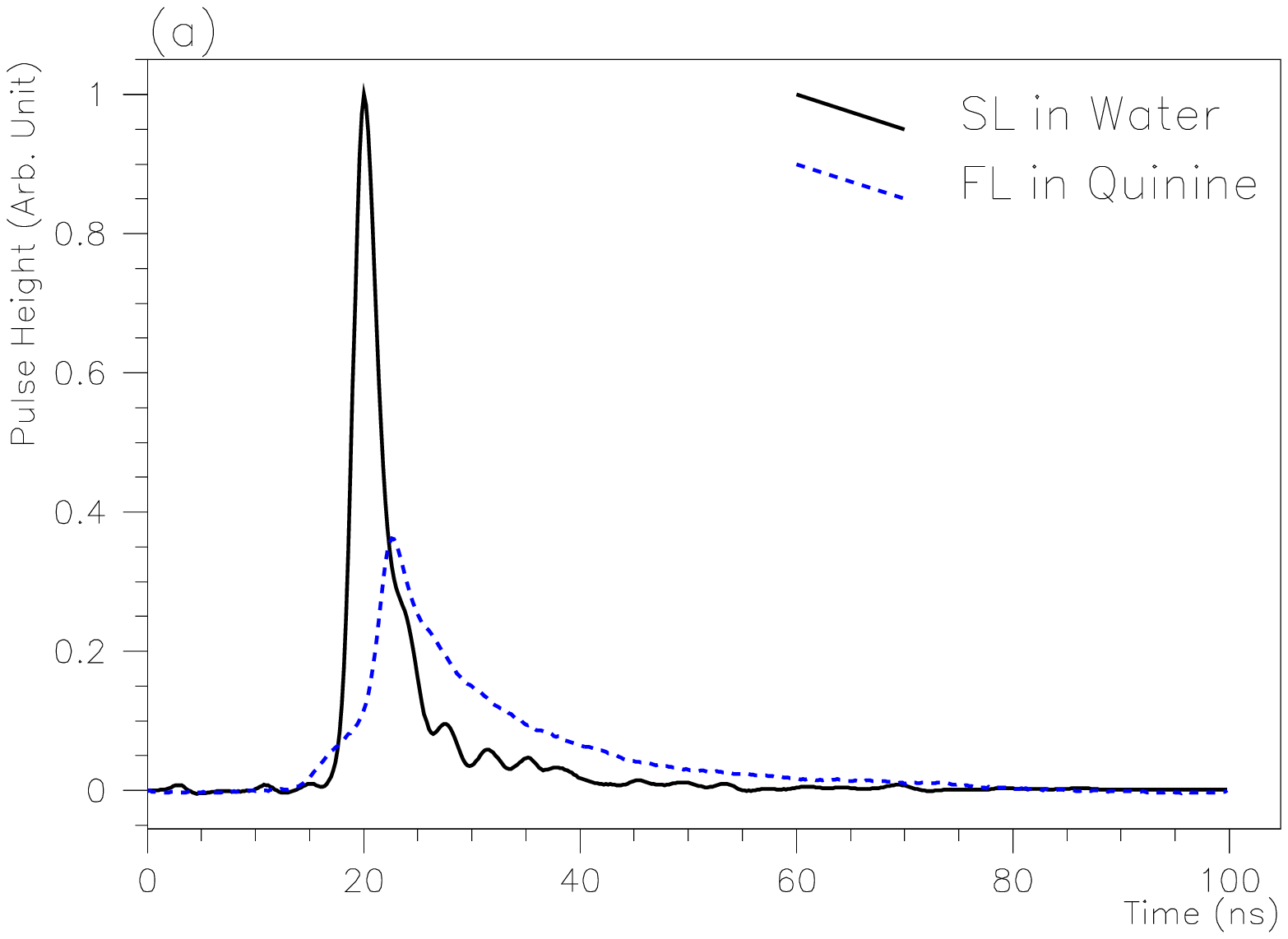,width=12cm}
}
\centerline{
\epsfig{figure=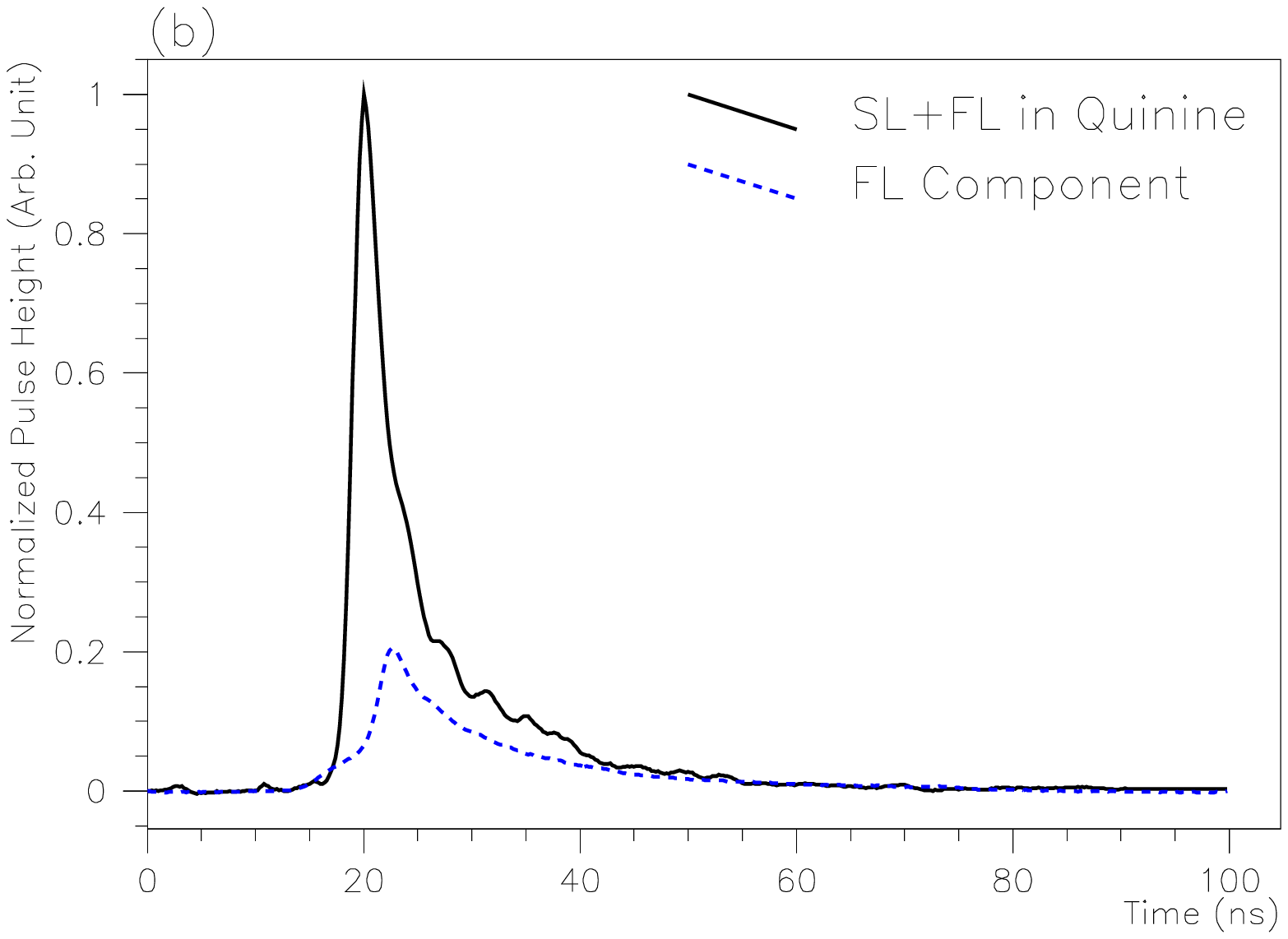,width=12cm}
}
\caption{
Measured pulse shapes of
(a) the reference SL in water and FL in
quinine-doped water due to cosmic-rays,
normalized to the same area, and
(b) combined SL and FL  emissions in
quinine-doped water, with the
deconvoluted FL component overlaid.
}
\label{slflpulse}
\end{figure}

The measured pulse shape for prompt SL emissions
in water ($\slop$)
is shown in Figure~\ref{slflpulse}a.
The SL emissions are expected to be extremely
fast at the 100~ps range\cite{gompf97,slwidth}, such that
the  measured shape of $\slop$ in
the ns scale reflects the response time 
of the PMT.

The pulse shape that corresponds to
pure FL emissions ($\flop$) is also displayed in
Figure~\ref{slflpulse}a.
It was obtained when the oscilloscope was
triggered by the CRT while quinine-doped 
water was placed in the flask.
Cosmic-ray events give rise
to FL emissions
in quinine-doped water,
due to (a) ionization and excitation
of the water molecules which transfer
the energy to the quinine molecules
leading to decays by fluorescence, 
and (b) absorption
of the Cherenkov light by the
quinine.
The measured pulse
shape is therefore dominated by
the FL process.
The decay time for $\flop$ is slower than
that for $\slop$, which is determined by
the energy transfer and
fluorescent processes in the quinine-doped water. 
Accordingly, the observed profile of $\flop$ in 
Figure~\ref{slflpulse}b is the actual
time-evolution of the FL emissions.
Both the SL- and CRT- trigger
PMTs provided timing to the oscilloscope
through the same electronic circuitry with 
cables of the same length, to synchronize 
the SL and FL reference pulses.

The observed pulse shape for SBSL emissions 
in quinine-doped water ($\pstot$), 
at the operating conditions of
10\% air in water and 
a temperature of 3$^o$C 
is shown in Figure~\ref{slflpulse}b.
Identical PMT and electronics triggering
schemes were adopted as the $\slop$ measurement
with pure water, so as to ensure good synchronization.
The pulse
consists of a fast component due to
SL emissions,
together with a slow component from
the secondary FL due to absorption
of the UV and high energy primary emissions.
Using an area normalization scheme:
$\rm{ 
\int \pstot (t) dt  = \int \slop (t) dt
= \int \flop (t) dt = 1 
  }$,
the various pulse profiles are related by:
\begin{displaymath}
\pstot (t) ~ = ~ [ 1 - \alpha ] \cdot  
\slop ( t + \delta _{SL} ) ~ + ~ 
\alpha \cdot \flop  ( t + \delta _{FL} ) ~. 
\end{displaymath}
The parameter $\alpha$ represents 
the relative strength of 
FL emissions to the total signal output,
while $\delta _{SL}$ and $\delta _{FL}$ 
denote possible residual 
time-shifts between the two reference pulses
with respect to the $\pstot$.

Fitting the measured
SBSL pulse shape in quinine-doped water
with a minimum-$\chi ^2$ algorithm 
in the ROOT software framework\cite{root},
the best-fit values  of
$\alpha$=$0.378 \pm 0.003 $,
$\delta _{SL}$=$( 0.014 \pm 0.005 )$~ns
and
$\delta _{FL}$=$( 0.094 \pm 0.044 )$~ns
at a
$\chi ^2$ per degree-of-freedom ($\chi ^2$/dof)
of 468/497 were derived.
Therefore, the FL component in 
SBSL under such conditions, 
as de-convoluted and depicted in
Figure~\ref{slflpulse}b, contributes
to about 38\% of the photo-electron yield
in the PMT, and can be measured to
$\sim 1\%$ accuracy.
The values of $\delta _{SL}$ 
and $\delta _{FL}$ are close to zero,
or alternatively, setting 
$\delta _{SL} = \delta _{FL} = 0$
leads only to a 1.3\% change in $\alpha$.
Both observations
indicate that the devised experimental procedures
did allow good time-synchronizations 
among $\pstot$, $\slop$ and $\flop$.

\section{Research Plans}

We have devised and demonstrated
a method to convert the UV photons
to observable optical signals which 
can be deconvoluted from
the primary SL emissions. 
This technique allows
quantitative studies on the emissions 
of high-energy radiations in Single-Bubble
as well as Multi-Bubble Sonoluminescence
which may shed lights on its underlying physics
mechanism.

After proper calibrations,
the measured $\alpha$-parameter 
can be translated to a ratio of energy 
output between UV and optical photons.
This can be used to test and differentiate
black-body and other theoretical
models in SBSL\cite{rmp02,slmodels}.
In the case of black-body spectra, 
the temperature 
at the light-emitting
surface can be obtained, and this
can be compared to
that derived from the measurement of the
optical spectra\cite{spectral}.

Systematic studies can be done with varying
ambient conditions like temperature, gas
and liquid compositions, 
acoustic pressure and driving frequency, 
to investigate
the relative changes in $\alpha$.
An increase in $\alpha$ would imply 
enhanced high energy emissions relative
to the optical light output.
In particular, in
the quest of identifying appropriate 
operating parameters which may favor
higher temperature emissions in SL,
a direct probe of the high energy radiations
will be extremely essential.

Towards these ends, research efforts are
pursued  to devise the calibration schemes
and to improve on the experimental hardware. 
In particular, a 2~GHz 4-channel
Sampling Analog Digital Convertor
running on VME-bus and Linux Operating System
will be installed to replace 
the oscilloscope readout. 
This will make efficient event-by-event
data taking with massive data flow possible.
Large data samples will allow the studies
of the $\alpha$-distribution,
from which one can investigate
if anomalous events may occur with
small but finite probabilities.

We are grateful to the referee for critical 
reading and comments. 
This work was supported by contracts
92-2112-M-001-057 and 
93-2112-M-001-030 
from the National Science Council, Taiwan,

\newpage

\newpage

\begin{thebibliography}{99}
\bibitem{slreview} 
B.R.~Barber et al., Phys. Report {\bf 281}, 65 (1977);\\
S.J.~Putterman and K.R.~Weninger,
Annu. Rev. Fluid Mech., {\bf 32}, 445 (2000).
\bibitem{rmp02}
M.P.~Brenner, S.~Hilgenfeldt and D.~Lohse,
Rev. Mod. Phys. {\bf 7}, 425 (2002).
\bibitem{spectral}
R. Hiller, S.J. Putterman, and B.P. Barber,
Phys. Rev. Lett. {\bf 69}, 1182 (1992).
\bibitem{gompf97}
B. Gompf et al., Phys. Rev. Lett. {\bf 79}, 1405 (1997).
\bibitem{ornl}
R.P. Taleyarkhan et al., Science 295, 1868 (2002);\\
R.P. Taleyarkhan et al., Phys. Rev. {\bf E 69}, 036109 (2004).
\bibitem{plasma}
D.J. Flannigan and K.S. Suslick, Nature {\bf 434}, 52 (2005).
\bibitem{slmhz}
C. Camara, S. Putherman and E. Kirilov,
Phys. Rev. Lett. {\bf 92}, 124301 (2004).
\bibitem{surfactant}
F. Grieser and M. Ashokkumar, Adv. Colloid Interface Sci., {\bf 89-90},
423 (2001).
\bibitem{quininespectra}
H. Du et al., Photochemistry and Photobiology {\bf 68},
141 (1998); also at 
http://omlc.ogi.edu/spectra/PhotochemCAD/html/quinine(0.5M\_H2SO4).html .
\bibitem{waterabsorption}
T.I.~Quickenden and J.A.~Irvin, J. Chem. Phys. {\bf 72}, 4416 (1980);\\
R.M.~Pope and E.S.~Fry, Appl. Optics {\bf 36}, 8710 (1997).
\bibitem{slwidth}
M.J. Morgan et al., Nucl. Instrum. Methods {\bf B 96}, 651 (1995);\\
R. Pecha et al., Phys. Rev. Lett. {\bf 81}, 717 (1998).
\bibitem{root}
R. Brun and F. Rademakers, 
Nucl. Instrum. Methods {\bf A 389}, 81 (1997).
\bibitem{slmodels}
Y. An, these Proceedings.
\end{thebibliography}
\end {document}